\documentclass[10pt,conference]{IEEEtran}
\usepackage{epsfig,rotating,setspace,latexsym,amsmath,epsf,amssymb,amsfonts,bm,theorem,cite,enumerate,longtable,accents,url}
\usepackage{algorithm,algorithmic,graphicx,epsf,authblk,epstopdf,url,color,multirow,longtable}
\usepackage{mathtools}

\newtheorem{theorem}{Theorem}

\newtheorem{remark}{Remark}

\newenvironment{Proof}[1]{\medskip\par\noindent{\bf Proof:\,}\,#1}{{\mbox{\,$\blacksquare$}\par}}

\IEEEoverridecommandlockouts
\allowdisplaybreaks

\title{Digital Blind Box: Random Symmetric \\ Private Information Retrieval}

\author{Zhusheng Wang \qquad Sennur Ulukus\\
	\normalsize Department of Electrical and Computer Engineering\\
	\normalsize University of Maryland, College Park, MD 20742\\
	\normalsize  \emph{zhusheng@umd.edu} \qquad \emph{ulukus@umd.edu}}

\begin{document}

\maketitle

\begin{abstract}
We introduce the problem of random symmetric private information retrieval (RSPIR). In canonical PIR, a user downloads a message out of $K$ messages from $N$ non-colluding and replicated databases in such a way that no database can know which message the user has downloaded (user privacy). In SPIR, the privacy is symmetric, in that, not only that the databases cannot know which message the user has downloaded, the user itself cannot learn anything further than the particular message it has downloaded (database privacy). In RSPIR, different from SPIR, the user does not have an input to the databases, i.e., the user does not pick a specific message to download, instead is content with any one of the messages. In RSPIR, the databases need to send symbols to the user in such a way that the user is guaranteed to download a message correctly (random reliability), the databases do not know which message the user has received (user privacy), and the user does not learn anything further than the one message it has received (database privacy). This is the digital version of a blind box, also known as gachapon, which implements the above specified setting with physical objects for entertainment. This is also the blind version of $1$-out-of-$K$ oblivious transfer (OT), an important cryptographic primitive. We study the information-theoretic capacity of RSPIR for the case of $N=2$ databases. We determine its exact capacity for the cases of $K = 2, 3, 4$ messages. While we provide a general achievable scheme that is applicable to any number of messages, the capacity for $K\geq 5$ remains open.
\end{abstract}

\section{Introduction}

Gachapon is a vending machine-dispensed capsule toy by means of a roulette mechanism, which makes it random and unpredictable for customers\cite{gashapon_wiki}. In addition, gachapon is being adapted as a random-type item in online games and 3D printing, and its digital form is catching on quickly in the worldwide market \cite{Game_APP,3D_APP}. Due to packaging requirements prior to official distribution, gachapon is also referred to as a \emph{blind box} \cite{gashapon_wiki}. A blind box is a type of packaging that keeps its contents hidden. The covers of blind boxes are identical in every way. Nobody including the manufacturer knows what exactly is inside until the customer opens a blind box\cite{blindbox_webdefinition}. Nowadays, not only constrained to the scope of entertainment, blind box has become a commercial phenomenon in certain parts of the world impacting people's daily lives.

Following the concepts of gachapon as well as blind box, we introduce a \emph{digital blind box} between a user and a server in a communication network with the following characteristics: 1) A user will ultimately receive a random box (content) from the server. However, the user does not know anything about what is in the box (what the content is) until it receives a box (content) from the server. 2) For the sake of unpredictability, a user should also know nothing about the current box (content) based on what it has received in the previous transactions. A user should not know anything about what other users have received before communicating with them. In other words, a user should not know anything beyond what it receives from the current box (content). This requirement also protects the content privacy of the server. 3) In order to protect the privacy of the users, the server should learn nothing about what a specific user has received.

Introduced in \cite{PIR, PIR_ORI}, private information retrieval (PIR) characterizes a fundamental problem, where a user downloads a message out of multiple messages stored in several non-colluding and replicated databases in such a way that no single database can know which message the user has downloaded. This privacy requirement is referred to as \emph{user privacy}. Some important variations of the PIR problem have been investigated in \cite{MMPIR,one_extra_bit,PIR_coded,Kumar_PIRarbCoded,ChaoTian_coded_minsize,PIR_WTC_II,PIR_cache_edge,ChaoTian_leakage,WeaklyPIR,Tian_upload,securestoragePIR,BPIRjournal,XSTPIR,XSTPIR_MDS,tandon-attia,HeteroPIR,Tamo_journal,AsymmetryHurtsPIR,PrivateComputation,NoisyPIR,PrivateSearch,SemanticPIR,SDB_PIR,SDB_PIR_LRC,SDB_MMPIR1,SDB_MMPIR2,tandon_cache_2017,Cache-aided_PIR,PrefetchingPIR,PartialPSI_PIR,StorageConstrainedPIR_Wei,PIR_PSI,MMPIR_PrivateSideInfo}. Further extended in \cite{SPIR,SPIR_ORI}, symmetric PIR (SPIR) requires in addition that the user learns nothing about the remaining messages stored in the databases after downloading its desired message. This privacy requirement is referred to as \emph{database privacy}. Some important variations of SPIR problem have been investigated in \cite{PSI_journal,MP-PSI_journal,SPIR_atPIR,Min_Uploadcost_SPIR,CommCost_ISIT2022,SPIR_Eavesdropper,SPIR_Mismatched,SPIR_coded,BlindSPIR,SPIR_Collusion}. 

In this paper, we introduce a new concept called random SPIR (RSPIR). In reference to the conventional SPIR, the only difference is that, in RSPIR there is no input at the user side. That is, the user does not send any queries to the databases, and ultimately receives a random message from the databases. This requirement is referred to as \emph{random reliability}. Interestingly, the three requirements of RSPIR, namely, random reliability, database privacy and user privacy, strictly correspond to the three characteristics of the digital blind box described above. Thus, the digital blind box is equivalent to the RSPIR.

Oblivious transfer (OT), first introduced in \cite{OT} and then developed in \cite{ChosenOT}, is an essential building block in modern cryptography. A $1$-out-of-$K$ OT protocol consists of two parties, a sender with $K$ input messages and a receiver with a choice $k \in [K]$. The objective of the protocol is that the receiver will receive the $k$th message without the sender learning the index $k$, while the sender can guarantee that the receiver only received one of the $K$ messages. Note that SPIR is a distributed (multi-database) version of $1$-out-of-$K$ OT. An important variant of $1$-out-of-$K$ OT is that the receiver has no input. Thus, the receiver will receive each potential message with equal probability without gaining any partial knowledge about the remaining messages, while the sender is ignorant of which message has been received by the receiver. For example, this variant can be used as a subroutine in contract signing and certified mail protocols \cite{ChosenOT}. Likewise, RSPIR can be viewed as a distributed version of this variant of $1$-out-of-$K$ OT.

Another instance of RSPIR can be observed in the problem formulation in \cite{SPIR_atPIR} which considers the SPIR problem with user-side common randomness. The problem formulation in \cite{SPIR_atPIR} allows the user to fetch a random subset of the common randomness available at the databases to form user-side side-information unknown to the databases (unknown also to the user before it receives them). The purpose of this action is to increase the SPIR rate; in fact, such an action increases the SPIR rate to the level of PIR rate. The common randomness fetching phase of \cite{SPIR_atPIR} is an RSPIR problem. 

In this paper, we formulate $N=2$ database RSPIR and investigate its capacity. We determine its capacity as well as the minimal amount of required common randomness in the cases of $K = 2, 3, 4$ messages. This determines the capacity of digital blind box. While we give a general achievable scheme for any number of messages, the exact capacity of RSPIR for $K \geq 5$ remains an open problem. 

\section{RSPIR: Problem Formulation}
In this paper, we consider $N = 2$ non-colluding databases each storing the same set of $K \geq 2$ i.i.d.~messages. Each message consists of $L$ i.i.d.~uniformly chosen symbols from a sufficiently large finite field $\mathbb{F}_q$, i.e.,
\begin{align}
H(W_{k}) &= L, \quad k \in [K]\\
H(W_{1:K}) &= H(W_1) + \cdots + H(W_K) = KL
\end{align}

The two databases jointly share a necessary common randomness random variable $\mathcal{S}$, which is generated independent of the message set $W_{1:K}$. Thus,
\begin{align}
H(W_{1:K},\mathcal{S}) = H(W_{1:K}) + H(\mathcal{S}) \label{message and randomness independence}
\end{align}

Before the RSPIR process starts, an answer set $\mathcal{A}$ with cardinality $M_1$ is given to database $1$, and another answer set $\mathcal{B}$ with cardinality $M_2$ is given to database $2$. Because of the fact that there is no input at the user side in the RSPIR process, the databases will never receive a query from the user. Therefore, both databases will independently select a random answer under a uniform distribution from their corresponding answer sets and then send them to the user. The indices of the answers for two databases are denoted by $a$ and $b$, respectively, i.e., database $1$ will select $A_a \in \mathcal{A}$ and database $2$ will select $B_b \in \mathcal{B}$. We note that every answer from any answer set is generated based on the message set and the common randomness, hence, for all $a \in [M_1]$ and $b \in [M_2]$, we have
\begin{align}
 \text{[deterministic answer]} \quad H(A_a, B_b|W_{1:K},\mathcal{S}) = 0 \label{determined answer}
\end{align}

After collecting two arbitrary answers from the databases, the user should always be able to decode a random message reliably. Thus, for all $a \in [M_1]$ and $b \in [M_2]$, we can always find an index $\theta_{a,b} \in [K]$ such that 
\begin{align}
\text{[random reliability]} \quad H(W_{\theta_{a,b}}|A_a,B_b) = 0 \label{random reliability}
\end{align}

Because of the database privacy constraint, the user is supposed to learn nothing about $W_{\bar{\theta}_{a,b}}$ which is the complement of the random decodable message $W_{\theta_{a,b}}$, i.e., $W_{\bar{\theta}_{a,b}} = \{W_1,\dots,W_{\theta_{a,b}-1},W_{\theta_{a,b}+1},\dots,W_K\}$, 
\begin{align}
\text{[database privacy]} \quad I(W_{\bar{\theta}_{a,b}};A_a,B_b) = 0 \label{database privacy}
\end{align}

Because of the user privacy constraint, i.e., the protection of the retrieved random message index of the user, from the perspective of each individual database, this index must be indistinguishable for each randomly selected answer under a uniform distribution. In other words, even though an answer from one database is deterministic, the user can still decode every potential message in the message set with equal probability through the variation of the answer from the other database. Thus, for any realization $a \in [M_1]$, we always have the following probability distribution of the random variable $\theta_{a,b}$ with respect to $b$,
\begin{align}
P(\theta_{a,b} = k) = \frac{1}{K}, \quad \forall k \in [K]
\label{user privacy 1}
\end{align}
which is equivalent to
\begin{align}
\text{[user privacy]} \quad I(a,A_a,W_{1:K},\mathcal{S};\theta_{a,b}) = 0
\label{user privacy 2}
\end{align}

By symmetry, for any realization $b \in [M_2]$, we also have the following probability distribution of the random variable $\theta_{a,b}$ with respect to $a$,
\begin{align}
P(\theta_{a,b} = k) = \frac{1}{K}, \quad \forall k \in [K]
\label{user privacy 3}
\end{align}
which is equivalent to
\begin{align}
\text{[user privacy]} \quad I(b,B_b,W_{1:K},\mathcal{S};\theta_{a,b}) = 0
\label{user privacy 4}
\end{align}

As a consequence, we obtain the following theorem regarding the cardinality of the answer sets, which can be proved by contradiction using the user privacy constraint. 

\begin{theorem}
    The total possible number of answers in the answer set for each database must be a multiple of $K$, i.e.,
    \begin{align}
        M_1 = t_1K,  ~ M_2 = t_2K, \quad t_1, t_2 \in N^+
    \end{align}
\end{theorem} 

Moreover, we also have the following theorem concerning the common randomness distribution in the databases.

\begin{theorem}
As in multi-database SPIR \cite{SPIR_ORI,SPIR}, in RSPIR, the databases must share some necessary common randomness that is unknown to the user before the retrieval process starts. Otherwise, RSPIR is not feasible.
\end{theorem}

\begin{Proof}
Without the existence of common randomness in the databases, for any $a \in [M_1]$ and $b \in [M_2]$, we always have the random reliability constraint $H(W_{\theta_{a,b}}|A_a,B_b) = 0$ and the database privacy constraint $I(W_{\bar{\theta}_{a,b}};A_a,B_b) = 0$, which lead to
\begin{align}
    0 &= I(W_{\bar{\theta}_{a,b}};A_a,B_b) \\
    &= I(W_{\bar{\theta}_{a,b}};W_{\theta_{a,b}},A_a,B_b) \\ 
    &= H(W_{\bar{\theta}_{a,b}}) - H(W_{\bar{\theta}_{a,b}}|W_{\theta_{a,b}},A_a,B_b) \label{thm2.1}
\end{align}
Then, we consider the following expression
\begin{align}
    &I(A_a,B_b;W_{\bar{\theta}_{a,b}}|W_{\theta_{a,b}}) \notag \\
    &\quad = H(W_{\bar{\theta}_{a,b}}|W_{\theta_{a,b}}) - H(W_{\bar{\theta}_{a,b}}|W_{\theta_{a,b}},A_a,B_b) \\
    &\quad = H(W_{\bar{\theta}_{a,b}}) - H(W_{\bar{\theta}_{a,b}}) \label{thm2.2} \\
    &\quad = 0 \label{thm2.3}
\end{align}
where \eqref{thm2.2} follows from \eqref{thm2.1}. For any arbitrary fixed $a$, 
\begin{align}
    0 &= I(A_a;W_{\bar{\theta}_{a,b}}|W_{\theta_{a,b}}) \label{thm2.4}\\
    &= H(A_a|W_{\theta_{a,b}}) -  H(A_a|W_{1:K}) \\
    &= H(A_a|W_{\theta_{a,b}}) \label{theorem2.1}
\end{align}
where (\ref{thm2.4}) follows from (\ref{thm2.3}), and \eqref{theorem2.1} follows from $H(A_a|W_{1:K}) = 0$. Taking into consideration the fact that \eqref{theorem2.1} is true for any realization $b \in [M_2]$ as well as the user privacy constraint \eqref{user privacy 1}, we have $H(A_a|W_1) = \cdots = H(A_a|W_K) = 0$. Since messages are all mutually independent, it is easy to derive that $H(A_a) = 0$, which forms a contradiction.
\end{Proof}

A valid two-database RSPIR achievable scheme is a scheme that satisfies the user privacy constraint \eqref{user privacy 2}, \eqref{user privacy 4}, the database privacy constraint \eqref{database privacy} and the random reliability constraint \eqref{random reliability}. The efficiency of the scheme is measured in terms of the maximal number of downloaded bits by the user from two databases, named as download cost and denoted by $D_{RSPIR}$. Thus, the retrieval rate of RSPIR is given by
\begin{align} 
R_{RSPIR} = \frac{L}{D_{RSPIR}} \label{ratedefinition}  
\end{align}
The capacity of RSPIR, $C_{RSPIR}$, is the supremum of the retrieval rates $R_{RSPIR}$ over all valid achievable schemes.

\section{Main Results}
\begin{theorem} \label{main theorem 1}
In the two-database RSPIR problem, in the case of $K = 2$, the capacity is $\frac{1}{2}$ with minimal amount of required common randomness being $L$. In the case of $K = 3, 4$, the capacity is $\frac{1}{3}$ with minimal amount of required common randomness being $2L$.
\end{theorem}

The converse proof of Theorem~\ref{main theorem 1} is given in Section~\ref{Converse Proof}, and the achievability proof of Theorem~\ref{main theorem 1} is presented in Section~\ref{Achievability}. The capacity and its minimal amount of required common randomness in the case of $K \geq 5$ is an open problem.

\begin{remark}
It is well known \cite{SPIR} that the capacity of multi-database SPIR is $1 - \frac{1}{N}$, where $N$ is the number of replicated and non-colluding databases. As a corollary, the capacity of two-database SPIR is $\frac{1}{2}$, which does not depend on the number of messages $K$ stored in the databases. By contrast, the capacity of RSPIR does depend on the value of $K$. Even though the capacity of RSPIR achieves the same limit as SPIR in the case of $K = 2$, the capacity of RSPIR decreases to $\frac{1}{3}$ when the value of $K$ increases to $3$. 
\end{remark}

\begin{remark} 
Because of the equivalence between RSPIR and the digital blind box, in a digital blind box setting where two non-colluding databases share $K$ messages and some necessary common randomness, perfect digital blind box delivery can be achieved with a linear download cost $KL$. The proof is a direct consequence of the second general achievable scheme given in Section~\ref{Achievability}.  
\end{remark}

\begin{remark}
In the problem formulation part of our previous work \cite{SPIR_atPIR}, we assume that the user is able to obtain a random subset of the shared common randomness that is unknown to any individual database before the SPIR retrieval process starts. Although we mention the idea of fetching common randomness like side-information in advance, we do not specify in \cite{SPIR_atPIR} a corresponding practical implementation. Now, it is clear that the achievability provided here for RSPIR can be used as a practical approach for this problem if common randomness is treated as another independent message system. 
\end{remark}

\section{Converse Proof} \label{Converse Proof}
\begin{theorem} \label{theorem3}
In the two-database RSPIR problem, the capacity is realized in the case where $M_1$ and $M_2$ are both exactly $K$.
\end{theorem}
\begin{Proof}
We provide a sketch of proof here. The idea of the proof is that once we multiply the value of $M_1$ by an integer $t \geq 2$, it is straightforward to see that additional constraints will be added to each pair $A_a, B_b$ for all $a \in [tM_1]$ and $b \in [M_2]$ after considering the index permutation, which will either increase or maintain the minimal value of $H(A_a) + H(B_b)$. This analysis also applies to the increase of $M_2$.
\end{Proof}

In the case of $K = 2$, motivated by Theorem~\ref{theorem3}, we consider the simplest case where $M_1 = 2$ and $M_2 = 2$. Then, we only need to consider the following constraints since all the other potential system of constraints have the same structure as this one and will lead to the same conclusions,
\begin{align}
    &H(W_1|A_1,B_1) = 0 \label{pairwise0} \\
    &H(W_2|A_1,B_2) = 0 \\
    &H(W_2|A_2,B_1) = 0 \\
    &H(W_1|A_2,B_2) = 0 \label{pairwise1}
\end{align}
These constraints reflect the random reliability constraint \eqref{random reliability} and user privacy constraint \eqref{user privacy 1}, \eqref{user privacy 3} involved in this problem. First, we prove a lower bound for $H(A_1) + H(B_1)$,
\begin{align}
    H&(A_1) + H(B_1) \notag \\ 
    &\geq H(A_1|A_2,B_1) + H(B_1|A_2,B_2) \\
    &= H(A_1,A_2,B_1) + H(A_2,B_1,B_2) \notag \\
    &\quad - H(A_2,B_1) - H(A_2,B_2) \label{proof1.1} \\
    &= H(W_1,A_1,A_2,B_1) + H(W_1,A_2,B_1,B_2) \notag \\
    &\quad - H(A_2,B_1) - H(A_2,B_2) \\
    &\geq H(W_1,A_2,B_1) + H(W_1,A_1,A_2,B_1,B_2) \notag \\
    &\quad - H(A_2,B_1) - H(A_2,B_2) \\
    &= H(A_1,A_2,B_1,B_2) - H(A_2,B_2) + H(W_1) \label{proof1.2} \\
    &\geq H(W_2,A_2,B_2) - H(A_2,B_2) + H(W_1) \\
    &= H(W_2) + H(W_1) \label{proof1.3} \\
    &= 2L 
\end{align}
where \eqref{proof1.2} and \eqref{proof1.3} follow from the database privacy constraint. Likewise, we can always obtain $H(A_a) + H(B_b) \geq 2L$ for any other answer pair $A_a, B_b, a,b \in [2]$. As a result, we reach a converse result for the capacity when $K = 2$,
\begin{align} \label{converseK=2}
    R = \frac{L}{D} \leq \frac{L}{\max\limits_{a,b} H(A_a) + H(B_b)} \leq \frac{L}{2L} = \frac{1}{2}
\end{align}

Next, we prove the minimal required amount of common randomness shared in the two databases.
\begin{align}
    0 
    &= I(W_2;A_1,B_1) \\
    &= I(W_2;A_1,B_1|W_1) \\
    &= H(A_1,B_1|W_1) - H(A_1,B_1|W_1,W_2) \notag \\
    &\quad + H(A_1,B_1|W_1,W_2,\mathcal{S}) \label{proof2.1} \\
    &= H(A_1,B_1|W_1) - I(A_1,B_1;\mathcal{S}|W_1,W_2) \\
    &= H(A_1,B_1|W_1) - H(\mathcal{S}|W_1,W_2) \notag \\
    &\quad + H(\mathcal{S}|W_1,W_2,A_1,B_1) \\
    &\geq H(A_1,B_1|W_1) - H(\mathcal{S}) \label{proof2.2}
\end{align}
where \eqref{proof2.1} follows from the deterministic answer constraint \eqref{determined answer} and \eqref{proof2.2} follows from the independence between message set and the common randomness \eqref{message and randomness independence}. Therefore, we turn to find a lower bound for the expression $H(A_1,B_1|W_1)$,
\begin{align}
    H&(A_1,B_1|W_1) \notag \\
    &= H(A_1|W_1,B_1) + H(B_1|W_1) \\
    &\geq H(A_1|W_1,A_2,B_1) + H(B_1|W_1,A_2,B_2) \\
    &= H(A_1,A_2,B_1) + H(A_2,B_1,B_2) \notag \\
    &\quad - H(W_1,A_2,B_1) - H(A_2,B_2) \\
    &= H(A_1,A_2,B_1) + H(A_2,B_1,B_2) - H(A_2,B_1) \notag \\
    &\quad - H(A_2,B_2) - H(W_1) \label{proof3.1}\\
    &\geq H(W_2) \label{proof3.2} \\
    &= L 
\end{align}
where \eqref{proof3.1} follows from the database privacy constraint and \eqref{proof3.2} exactly follows from the steps between \eqref{proof1.1}-\eqref{proof1.3}. As a consequence, we reach a converse result for the minimal amount of required common randomness,
\begin{align}
    H(\mathcal{S}) \geq L
\end{align}

In the case of $K = 3$, $M_1$ and $M_2$ both take the value $3$, after converting the random reliability constraint and user privacy constraint into pairwise constraints as in \eqref{pairwise0}-\eqref{pairwise1}, we can proceed with the converse steps. As in the converse proof in the case of $K = 2$ above, the concrete process is to utilize the converse proof of \cite[Theorem~2]{CommCost_ISIT2022} once more after eliminating the influence of retrieval strategy randomness and its generated queries. Thus, we have the same conclusions as the one in \cite[Theorem~2]{CommCost_ISIT2022} in the case of $K = 3$,
\begin{align} \label{converseK=3}
    R \leq \frac{1}{3}, \quad H(\mathcal{S}) \geq 2L
\end{align}

In the case of $K = 4$, it is easy to verify that each answer pair $A_a, B_b, a,b \in [4]$ has more constraints than the one in the case of $K = 3$. Thus, a converse proof for the capacity and the minimal amount of required common randomness in the case of $K = 4$ can be inherited from the case of $K = 3$, i.e.,
\begin{align}\label{converseK=4}
    R \leq \frac{1}{3}, \quad H(\mathcal{S}) \geq 2L
\end{align}

A tight converse proof for the capacity and the minimal amount of required common randomness remains to be found in the case of $K \geq 5$.

\section{Achievability} \label{Achievability}
The work in \cite{Min_Uploadcost_SPIR} provides a scheme that can be readily converted into an achievable scheme (albeit suboptimal) for the  two-database RSPIR problem. For clarity, we restate the result from the new perspective of RSPIR here. Assuming that $L = 1$ for the time being, two databases share $K$ common randomness symbols $S_1, \dots, S_K$, which are all uniformly selected from a finite field $\mathbb{F}_q$. For database $1$, the answer set $\mathcal{A}$ is composed of $K$ elements in the following form,
\begin{align}
    A_1 &= (W_1 \!+\! S_1, W_2 \!+\! S_2, \dots, W_K \!+\! S_K) \\
    A_2 &= (W_1 \!+\! S_2, W_2 \!+\! S_3, \dots, W_K \!+\! S_1) \\
    \vspace{-0.25em}
    &\vdots \notag \\
    \vspace{-0.25em}
    A_K &= (W_1 \!+\! S_K, W_2 \!+\! S_1, \dots, W_K \!+\! S_{K-1}) 
\end{align}
Basically, we only rotate common randomness symbols by one place in the sequence of answers. A homomorphic variation of $\mathcal{A}$ is to rotate message symbols by one place without imposing any influence on the  answer set $\mathcal{B}$ and it is shown as follows,
\begin{align}
    A_1 &= (W_1 \!+\! S_1, W_2 \!+\! S_2, \dots, W_K \!+\! S_K) \\
    A_2 &= (W_2 \!+\! S_1, W_3 \!+\! S_2, \dots, W_1 \!+\! S_K) \\
    \vspace{-0.25em}
    &\vdots \notag \\
    \vspace{-0.25em}
    A_K &= (W_K \!+\! S_1, W_{1} \!+\! S_2, \dots, W_{K-1} \!+\! S_K) 
\end{align}
For database $2$, the answer set $\mathcal{B}$ also including $K$ elements is shown as follows,
\begin{align}
    B_1 &= S_1 \\
    B_2 &= S_2 \\
    \vspace{-0.25em}
    &\vdots \notag \\
    \vspace{-0.25em}
    B_K &= S_K 
\end{align}

The answer set construction in these two databases is public knowledge to the user. Afterwards, database $1$ selects a random answer under a uniform distribution from $\mathcal{A}$, and then sends the values of symbols as well as the index belonging to this answer to the user. Likewise, database $2$ performs the same selection and transmission. The reason for sending the answer indices is that the user does not know how to use the values of symbols in the answers to decode a random message without the help of the answer indices.  After receiving these two answers, the user is always able to decode one random message reliably. Moreover, since each database is doing the uniform selection, this random message is equally likely to be any message in the message set. Therefore, it is impossible for each individual database to learn the index of the retrieved random message at the user side. Meanwhile, the user cannot learn any information about the remaining messages because of the existence of unknown common randomness symbols. When each message includes multiple symbols, we can simply perform this scheme repeatedly for each symbol while there is no need to do the new selection nor send the answer index for each database after first execution. Thus, the download cost of answer indices can be ignored when $L$ is large enough. Obviously, the download cost is $D=(K+1)L$ in this scheme when $L$ approaches infinity, as database 1 answers have $K$ symbols each, and database 2 answers have 1 symbol each. However, this download cost is not optimal. 

Here, we provide a new general scheme that achieves the download cost of $D = KL$ when $L$ goes to infinity. Assuming that $L = 1$ temporarily, let the two databases share $K-1$ common randomness symbols $S_1, \dots, S_{K-1}$, which are all uniformly selected from a finite field $\mathbb{F}_q$. For database $1$, the answer set $\mathcal{A}$ contains $K$ elements in the following form,
\begin{align}
    A_1 &= (S_1,S_2, \dots, S_{K-1}) \\
    A_2 &= (W_1 \!+\! W_2 \!+\! S_1, W_2 \!+\! W_3 \!+\! S_2, \notag \\
    & \qquad \dots, W_{K-1} \!+\! W_K \!+\! S_{K-1}) \\
    A_3 &= (W_1 \!+\! W_3 \!+\! S_1, W_2 \!+\! W_4 \!+\! S_2, \notag \\
    & \qquad \dots, W_{K-1} \!+\! W_1 \!+\! S_{K-1}) \\
    \vspace{-0.25em}
    &\vdots \notag \\
    \vspace{-0.25em}
    A_K &= (W_1 \!+\! W_K \!+\! S_1, W_2 \!+\! W_1 \!+\! S_2, \notag \\
    & \qquad \dots, W_{K-1} \!+\! W_{K-2} \!+\! S_{K-1}) 
\end{align}
Basically, except for the first answer, we only rotate the second message symbol by one place in the sequence of answers while keeping the first message symbol. For database $2$, the answer set $\mathcal{B}$  consists of $K$ elements in the following form,
\begin{align}
    B_1 &= W_1 \!+\! S_1 \\
    B_2 &= W_2 \!+\! S_2 \\
    \vspace{-0.25em}
    &\vdots \notag \\
    \vspace{-0.25em}
    B_{K-1} &= W_{K-1} \!+\! S_{K-1}  \\
    B_K &= W_K \!+\! S_1 \!+\! S_2 \!+\! \dots S_{K-1}
\end{align}

Now, note that, since this scheme achieves a download cost of $D = KL$, it achieves a rate of $R=\frac{L}{D}=\frac{L}{KL}=\frac{1}{K}$. For $K=2$ and $K=3$, this scheme achieves rates $\frac{1}{2}$ and $\frac{1}{3}$ meeting the converse bounds in (\ref{converseK=2}) and (\ref{converseK=3}), respectively.

Specifically, when $K = 2$, we have the following answer sets that achieve the converse,
\begin{align}
    A_1 &= S_1 \quad B_1 = W_1 \!+\! S_1 \\
    A_2 &= W_1 \!+\! W_2 \!+\! S_2 \quad B_2 = W_2 \!+\! S_2
\end{align}

When $K=3$, we have the following answer sets that achieve the converse,
\begin{align}
    A_1 &\!=\! (S_1, S_2), \; B_1 \!=\! W_1\!+\!S_1 \label{ach1}\\
    A_2 &\!=\! (W_1\!+\!W_2\!+\!S_1, W_2\!+\!W_3\!+\!S_2), \; B_2 \!=\! W_2\!+\!S_2 \label{ach2} \\
    A_3 &\!=\! (W_1\!+\!W_3\!+\!S_1, W_2\!+\!W_1\!+\!S_2), \; B_3 \!=\! W_3\!+\!S_1\!+\!S_2 \!\!\! \label{ach3}
\end{align}

\begin{figure}[ht]
	\centering
	\includegraphics[width=0.48\columnwidth]{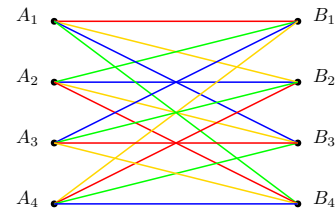}
	\caption{A two-database RSPIR bipartite graph for $K = 4$ messages.}
	\label{K4U4_Full}
\end{figure}

When $K=4$, this achievable scheme achieves a rate $R=\frac{1}{K}=\frac{1}{4}$ whereas the converse in (\ref{converseK=4}) gives a bound of $\frac{1}{3}$. Now, we provide a better scheme that achieves the converse in the case of $K = 4$. The message length $L$ is assumed to be $2$ such that $W_1 = \{a_1,a_2\}, W_2 = \{b_1,b_2\}, W_3 = \{c_1,c_2\}$ and $W_4 = \{d_1,d_2\}$. Moreover, two databases share $4$ common randomness symbols $S_1, S_2, S_3, S_4$, which are all uniformly selected from a finite field $\mathbb{F}_q$. For database $1$, the answer set $\mathcal{A}$ contains $4$ elements in the following form,
\begin{align}
    A_1 \!&=\! (S_1,S_2,S_3) \\
    A_2 \!&=\! (a_1\!+\!c_1\!+\!c_2\!+\!S_1,b_2\!+\!d_1\!+\!S_1\!+\!S_3,c_2\!+\!S_4) \\
    A_3 \!&=\! (a_1\!+\!d_2\!+\!S_1\!+\!S_4,a_2\!+\!d_1\!+\!d_2\!+\!S_2,b_1\!+\!c_2\!+\!S_2\!+\!S_3) \\
    A_4 \!&=\! (b_1\!+\!S_4,a_1\!+\!a_2\!+\!b_1\!+\!b_2\!+\!S_1\!+\!S_2,c_1\!+\!d_2\!+\!S_1\!+\!S_2\!+\!S_3)
\end{align}
For database $2$, the answer set $\mathcal{B}$ consists of $4$ elements in the following form,
\begin{align}
    B_1 \!&=\! (a_1\!+\!S_1,a_2\!+\!S_2,S_4) \\
    B_2 \!&=\! (b_1\!+\!b_2\!+\!S_1\!+\!S_2,b_1\!+\!S_2\!+\!S_3,a_1\!+\!c_1\!+\!d_2\!+\!S_1\!+\!S_4) \\
    B_3 \!&=\! (d_1\!+\!d_2\!+\!S_2,b_1\!+\!c_2\!+\!S_4,d_1\!+\!S_1\!+\!S_3) \\
    B_4 \!&=\! (c_2\!+\!S_2\!+\!S_3,c_1\!+\!c_2\!+\!S_1,a_1\!+\!a_2\!+\!b_2\!+\!c_1\!+\!d_1\!+\!S_3\!+\!S_4)
\end{align}
Here, the download cost is $D=6$ and the rate is $R=\frac{L}{D}=\frac{1}{3}$. The remaining steps and verification of this specific achievable scheme are the same as the last two general ones. Specifically, regarding verification, we can use the bipartite graph in Fig.~\ref{K4U4_Full}. In this bipartite graph, by using colors red, yellow, green and blue for messages $W_1, W_2, W_3$ and $W_4$, respectively, the color of links indicates which message should be decoded while keeping all the other messages private. Moreover, each node is always connected to $4$ links with different colors.

\newpage

\bibliographystyle{unsrt}
\bibliography{ITW2022}

\end{document}